\documentclass[conference]{IEEEtran}
\IEEEoverridecommandlockouts

\usepackage{amsmath,amssymb,amsfonts,bm}
\usepackage{graphicx}
\usepackage{subcaption}
\usepackage{cite}
\usepackage{siunitx}
\usepackage{algorithm}
\usepackage{algpseudocode}
\usepackage{standalone}   
\usepackage{pgfplots}

\usepackage{placeins}  

\tikzset{every picture/.style={/tikz/font=\footnotesize}}

\pgfplotsset{compat=newest}

\usetikzlibrary{positioning}
\usetikzlibrary{intersections}

\usepackage{tikz}
\usepackage[dvipsnames]{xcolor}
\usepackage{standalone}
\usetikzlibrary{arrows.meta,calc,positioning,decorations.pathreplacing}
\definecolor{tileBlue}{RGB}{150,180,220}
\definecolor{panelEdge}{RGB}{30,30,30}
\definecolor{dashBlue}{RGB}{45,95,160}
\definecolor{phonePurple}{RGB}{168,100,170}

\usepackage{tikz}
\tikzset{
  pics/phone/.style={code={
    \fill[black,rounded corners=0.5mm] (0,0) rectangle (0.6,1.1);
    \fill[rounded corners=0.5mm, white!90!black] (0.05,0.15) rectangle (0.55,0.95);
    \draw[white, thick] (0.22,1.025) -- (0.38,1.025);
    \fill[white](0.3,0.075) circle (.4ex);
  }}
}
            
\title{Optimizing Movable Antenna Position and Transmissive RIS Phase for Efficient Base Station Design}

\author{
\IEEEauthorblockN{Marjan Boloori, Chu Li, Aydin Sezgin}
\IEEEauthorblockA{Department of Digital Communication Systems, Ruhr University Bochum, Germany \\
Email: \{marjan.boloori, chu.li, aydin.sezgin\}@rub.de}
}

\begin{document}
\maketitle

\begin{abstract}
Movable antennas (MA) and transmissive reconfigurable intelligent surfaces (TRIS) represent two innovative technologies that significantly enhance the flexibility of wireless communication systems. 
In this paper, we propose a novel and compact base station architecture that synergistically integrates a movable antenna with a transmissive RIS in the near field, enabling joint optimization of antenna positioning and TRIS phase adjustments. The proposed model compensates for phase quantization loss and significantly enhances signal strength, even with low-resolution (1–2 bit) phase shifters. 
Leveraging this framework, we systematically evaluate system performance as a function of TRIS size and antenna placement. Our results indicate that antenna mobility provides an additional degree of freedom to enhance the desired signal and achieve a higher SNR, particularly when combined with TRIS capabilities. These findings demonstrate that MA–TRIS integration offers a cost-effective and energy-efficient pathway toward compact 6G base stations, combining hardware simplicity with strong performance gains.
\end{abstract}


\section{Introduction}
Reconfigurable intelligent surfaces (RIS) have emerged as a key enabling technology for sixth-generation (6G) wireless communication systems, as they can enhance network capacity and coverage by intelligently reconfiguring the wireless propagation environment~\cite{wu2019towards, liu2021reconfigurable}. An RIS consists of passive elements capable of manipulating propagation channels by adjusting the phase-shift and amplitude of incident electromagnetic waves. When deployed to assist communication, RIS can suppress interference and enhance desired signal power, thereby improving system performance. Furthermore, owing to its nearly passive full-duplex operation, RIS reflects or transmits incident signals without active amplification, thus avoiding self-interference and additional noise typically associated with conventional relay technologies~\cite{zuo2020resource}. Since RIS does not require complex signal processing modules, it exhibits low power consumption and reduced implementation cost, making it a strong candidate for future wireless networks~\cite{yan2020passive}.  

Recent studies classify RIS into two operational modes: reflective~\cite{li2021joint, chen2022irs} and transmissive~\cite{zeng2021reconfigurable, 10886969}. While the reflective mode has received significant attention for its ability to enhance system performance, the transmissive RIS (TRIS) mode is gaining interest for its potential to improve coverage and address blind spots~\cite{li2023toward}.
A closer examination of these two modes reveals that TRIS offers several advantages over its reflective counterpart. In reflective RIS configurations, the transmitter antenna and receiver are located on the same side of the surface, increasing the risk of self-interference through signal coupling. TRIS addresses this limitation by positioning the transmitter and receiver on opposite sides, which not only enables a more compact base station design but also significantly reduces interference. Additionally, TRIS supports wider operational bandwidths and higher aperture efficiency, positioning it as a versatile and high-performing solution for next-generation wireless networks~\cite{bai2020high, ali2025hybrid}.

Parallel to these advances, movable antennas (MA) have emerged as a promising technique for improving communication and sensing capabilities by enabling controlled movement of the radiating element, thereby introducing an additional spatial degree of freedom. Existing studies have explored the advantages of MAs for wireless communications. For example, \cite{zhu2023modeling}, \cite{zhu2024performance}, and \cite{mei2024movable}, demonstrated that employing MAs can efficiently enhance the received signal-to-noise ratio (SNR) based on deterministic or stochastic channel models \cite{10643473}.

Building on these insights, this study investigates a compact base station architecture that integrates a MA with a TRIS positioned within their near-field region. In scenarios where both the TRIS and MA exhibit pronounced spatial coupling, their joint deployment introduces new possibilities for significant performance enhancements~\cite{zhu2025tutorial}. The proposed configuration offers a streamlined and compact design, enabling strong SNR performance through optimal selection of TRIS dimensions and MA placement.

In practical implementations, TRIS hardware often employs discrete phase-shifters with limited resolution, typically in the range of 1–2 bits \cite{song2024modeling}. While phase quantization poses challenges for precise beam focusing, particularly in near-field scenarios, the system maintains remarkable flexibility for performance optimization. By leveraging the spatial degree of freedom through MA repositioning, the inherent performance loss from phase quantization can be effectively mitigated. This integration of spatial adaptability and hardware-efficient TRIS design not only ensures robust operation with low-resolution phase control but also facilitates cost-effective deployment strategies. In fact, MA repositioning introduces crucial spatial diversity that can offset the limitations of phase quantization, allowing the system to maintain robust performance and, in favorable scenarios, even approach or surpass the performance of ideal continuous phase-shifter implementations. To the best of our knowledge, this promising approach has not been systematically explored in prior literature, highlighting its novelty and practical relevance, and offering a positive outlook for future wireless communication systems.

\section{System Model}

\begin{figure}[t]
\centering

    \begin{tikzpicture}[
            x=5mm,y=5mm,
            yslant=0.18,               
            scale=0.8,
            >={Latex[length=1mm]},
             every path/.style={line width=0.15pt}, 
            ray/.style={line width=0.5pt},
            label/.style={scale=0.9,inner sep=1pt}
          ]
        
          \begin{scope}[yslant=-0.18]
            \draw[dashBlue, line width=1.2pt, dashed, dash pattern=on 7pt off 5pt, rounded corners=15pt]
              (-3.7,-1.0) rectangle (10.2,11);
            \node[dashBlue, font=\small] at (3.25,-2) {Base Station};
          \end{scope}
        
          \begin{scope}[shift={(1.5,0)}]
          
          \def\W{8}
          \def\N{6}
          \def\s{0.80}     
          \def\g{0.52}     
          \pgfmathsetmacro{\m}{(\W - (\N*\s + (\N-1)*\g))/2}
          \coordinate (C) at ({\W/2},{\W/2}); 
        
          \fill[white] (0,0) rectangle (\W,\W);
          \draw[panelEdge, line width=1pt] (0,0) rectangle (\W,\W);               

          \foreach \i in {0,...,5}{
            \foreach \j in {0,...,5}{
              \pgfmathsetmacro\x{\m + \i*(\s+\g)}
              \pgfmathsetmacro\y{\m + \j*(\s+\g)}
              \fill[ tileBlue!40!white ] (\x,\y) rectangle ++(\s,\s);
              \draw[panelEdge!70, line width=0.5pt] (\x,\y) rectangle ++(\s,\s);
            }
          }

            \coordinate (RIS_bl) at (0,0);
            \coordinate (RIS_tl) at (0,\W);
            \coordinate (RIS_br) at (\W,0);
            \coordinate (RIS_tr) at (\W,\W);

            \end{scope}
            

            \begin{scope}[shift={(0,-0.2)}, yslant=0]
              \coordinate (MA) at (-2.2,7.1);
            
              \def\boxsize{1.6} 
              \draw[panelEdge, line width=0.5pt, dashed, rounded corners=3.5pt]
                ($(MA)+(-\boxsize/2,-\boxsize/2)$) rectangle
                ($(MA)+(\boxsize/2,\boxsize/2)$);
            \node[anchor=north, label] at ($(MA)+(0,2)$) {$a$};
            \node[anchor=north, label] at ($(MA)+(1.5,0)$) {$a$};
            
              \draw[panelEdge!80, line width=1.0pt] (MA) -- ++(0,-0.4);
            
                \path let \p1=(MA) in
                  coordinate (B) at (\x1-4.0,\y1+4)        
                  coordinate (D) at (\x1+4.0,\y1+4)        
                  coordinate (A) at (\x1,\y1-1.20);       
                \draw[fill=tileBlue!60, line width=0.5pt] (B) -- (D) -- (A) -- cycle;

                \path
                  ($(MA)+(-\boxsize/2,\boxsize/2)$)  coordinate (MA_tl)
                  ($(MA)+(\boxsize/2,\boxsize/2)$)   coordinate (MA_tr)
                  ($(MA)+(-\boxsize/2,-\boxsize/2)$) coordinate (MA_bl)
                  ($(MA)+(\boxsize/2,-\boxsize/2)$)  coordinate (MA_br);
                \draw[panelEdge, line width=0.5pt, dashed, rounded corners=3.5pt]
                  (MA_bl) rectangle (MA_tr);

            \draw[panelEdge!80, <->, line width=0.55pt]
              ($(MA)+(-1,1)$) -- ($(MA)+(1,1)$);
            
            \draw[panelEdge!80, <->, line width=0.55pt]
              ($(MA)+(1,-0.85)$) -- ($(MA)+(1,1)$);
        
              \node[anchor=north, label] at ($(MA)+(0.2,-1.3)$) {Movable};
              \node[anchor=north, label] at ($(MA)+(0.2,-1.9)$) {Antenna};
            \end{scope}

            \draw[panelEdge!25, line width=0.25pt] (MA_tl) -- (RIS_tl);
            \draw[panelEdge!25, line width=0.25pt] (MA_tr) -- (RIS_tr);
            \draw[panelEdge!25, line width=0.25pt] (MA_bl) -- (RIS_bl);
            \draw[panelEdge!25, line width=0.25pt] (MA_br) -- (RIS_br);
        
            \node[transform shape=true] at (16.5,-1.1) {%
              \tikz[baseline]{
                \pic[scale=1.6] {phone}; 
              }%
            };
            \node[below=12pt, font=\small] at (16.5, -1.1) {User};

            \draw[ray, -{Latex[length=2.6mm]}]
          (C) -- (16,-0.9)
          node[pos=0.75, below=-10pt, label] {$g$};
        
            \fill[panelEdge] (C) circle (0.06);
        
          \node[anchor=north west, label] at (3.5,-0.1) {Transmissive RIS};

          \draw[ panelEdge!80, line width=0.7pt, -{Latex[length=2.6mm]}]
          (-2.2,6.5) -- (C)
          node[pos=0.5, above right=1.5pt, label, black] {$f(\boldsymbol{t})$};
          
            
            
            

    \end{tikzpicture}
    
    \caption{System model, the suggested compact BS with TRIS and MA, and the user device at the far field.}
    \label{fig:SysModel}
\end{figure}
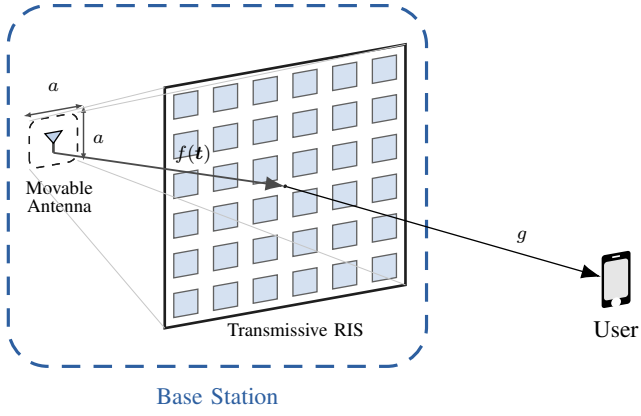

As shown in Fig. \ref{fig:SysModel}, we consider a compact base station (BS) equipped with a single MA and a TRIS. The MA is located at position $\boldsymbol{t} = [x_t, y_t, z_t]^T$ and can be flexibly moved within a two dimensional region of $\mathcal{R} = a \times a \, \mathrm{m^2}$.  

The TRIS consists of $N = N_\text{x} \times N_\text{y}$ transmissive elements, where $N_\text{x}$ and $N_\text{y}$ denote the number of elements arranged along the x- and y-axes, respectively. The TRIS is deployed in the near-field region of the MA and is assumed to be \emph{fully transmissive}, i.e., it does not reflect any incident signal \cite{10242373}.  

The Cartesian coordinates of the $(n_\text{x},n_\text{y})$-th TRIS element are given by $e_{n_\text{x}n_\text{y}} = [n_\text{x} d,\, n_\text{y} d,\, 0]^T$, where $d$ denotes the inter-element spacing. The index sets of the TRIS elements are defined as follows:
\begin{align}
    n_\text{x} &\in \{0, \pm 1, \pm 2, \dots, \pm N_\text{x}/2\}, \\
    n_\text{y} &\in \{0, \pm 1, \pm 2, \dots, \pm N_\text{y}/2\}.
\end{align}
The phase-shift configuration of the TRIS is represented by the coefficient vector
\begin{equation}
    \boldsymbol{\phi} = [e^{j\phi_1},\, e^{j\phi_2},\, \dots,\, e^{j{\phi}_{n_\text{x}n_\text{y}}},\,\dots, e^{j\phi_N}]^T ,
\end{equation}
where \(\phi_{n_\text{x}n_\text{y}} \in [0,2\pi)\) denotes the phase-shift applied to the $({n_\text{x},n_\text{y}})$-th transmissive unit.  

The near-field boundary, given by the Rayleigh distance, can be calculated as 
\begin{equation}
    D_\text{Ray} = 2D_\text{Aper}^2/\lambda
\end{equation}
where $\lambda$ represents the wavelength and $D_\text{Aper}$ denotes the largest dimension of the TRIS~\cite{RISNearField}. In this context, $D_\text{Aper} = N_\text{x}d_\text{x}$. This distance represents the spatial separation between the MA region and the TRIS. Subsequently, for each TRIS size, the relative Rayleigh distance for the MA is evaluated to ensure that both the MA and TRIS remain within the near-field region.

Let $f_{n_\text{x}n_\text{y}}$ denote the near-field channel coefficient between the MA and the TRIS, which is a function of MA position. The propagation distance between the MA and the \((n_\text{x},n_\text{y})\)-th TRIS element is given by
\begin{equation}
    d_{n_\text{x}n_\text{y}}^{\text{T}}
    = \left\| \boldsymbol{t} - e_{n_\text{x}n_\text{y}} \right\|_2,
\end{equation}
and accordingly, the channel coefficient can be expressed as
\begin{equation}\label{eq:f}
    f_{n_\text{x}n_\text{y}} = \sqrt{\frac{\lambda G_f F_f}{4\pi}} \frac{1}{d_{n_\text{x}n_\text{y}}^\text{T}}e^{-j\frac{2\pi}{\lambda}d_{n_\text{x}n_\text{y}}^\text{T}},
\end{equation}
where $G_f$ and $ F_f$ denote the antenna gain and the normalized power radiation pattern of the TRIS side facing the MA \cite{channelmodelFG}.
Similarly, the channel between the \((n_\text{x},n_\text{y})\)-th TRIS element and the user is denoted by $g_{n_\text{x}n_\text{y}}$.
Let \(d_{n_\text{x}n_\text{y}}^{\text{R}}\) be the propagation distance from the \((n_\text{x},n_\text{y})\)-th TRIS element to the user. Then, the channel coefficient is given by
\begin{equation}\label{eq:g}
    g_{n_\text{x}n_\text{y}} = \sqrt{\frac{\lambda G_g F_g}{4\pi}} \frac{1}{d_{n_\text{x}n_\text{y}}^\text{R}}e^{-j\frac{2\pi}{\lambda}d_{n_\text{x}n_\text{y}}^\text{R}},
\end{equation}
in which $G_g$ and $F_g$ correspond to the antenna gain and the normalized power radiation pattern of the TRIS side facing the user \cite{channelmodelFG}. 

Based on (\ref{eq:f}) and (\ref{eq:g}), the signal received by the user can be modeled as
\begin{align}\label{eq:ReceivedSignal}
    & y = \sqrt{P} \sum_{n_\text{x} = 1}^{N_\text{x}} \sum_{n_\text{y} = 1}^{N_\text{y}} f_{n_\text{x}n_\text{y}} g_{n_\text{x}n_\text{y}} \Gamma e^{j\phi_{n_\text{x}n_\text{y}}} s + n_0  \\ \notag
    & = \frac{\lambda \Gamma \sqrt{P G F }}{4\pi} \sum_{n_\text{x} = 1}^{N_\text{x}} \sum_{n_\text{y} = 1}^{N_\text{y}} \frac{1}{d_{n_\text{x}n_\text{y}}^\text{T}d_{n_\text{x}n_\text{y}}^\text{R}} e^{j ( \phi_{n_\text{x}n_\text{y}} - \frac{2\pi}{\lambda}(d_{n_\text{x}n_\text{y}}^\text{T}+d_{n_\text{x}n_\text{y}}^\text{R}))}s\\ \notag & + n_0,
\end{align}
where $P$ represents the transmit power, $n_0 \sim \mathcal{CN}(0, \sigma ^2)$ denotes the additive Gaussian noise with variance $\sigma ^2$, $s$ is the transmit signal, $\Gamma \in [0,1]$ is the transmission loss of the ${n_\text{x}n_\text{y}}$-th TRIS antenna and $G = G_fG_g$, and $F = F_fF_g$ \cite{channelmodelFG}.
The corresponding received SNR is
\begin{align} \label{eq:SNRFirst}
    & \text{SNR} = \\ \notag
    & \frac{\lambda ^2 \Gamma^2 {P G F }}{16\pi^2 \sigma ^2} \left | \sum_{n_\text{x} = 1}^{N_\text{x}} \sum_{n_\text{y} = 1}^{N_\text{y}} \frac{1}{d_{n_\text{x}n_\text{y}}^\text{T}d_{n_\text{x}n_\text{y}}^\text{R}} e^{j ( \phi_{n_\text{x}n_\text{y}} - \frac{2\pi}{\lambda}(d_{n_\text{x}n_\text{y}}^\text{T}+d_{n_\text{x}n_\text{y}}^\text{R}))}  \right |^2.
\end{align}

\section{Problem Formulation and Proposed Algorithm}

To assess the impact of MA, we evaluate (\ref{eq:SNRFirst}) within the frameworks of both discrete and continuous TRIS configurations. In scenarios where continuous phase-shifters are employed, each TRIS element is capable of achieving optimal phase alignment with the intended transmission direction, thereby eliminating any residual phase mismatch. It means that for each value of $\frac{2\pi}{\lambda}(d_{n_\text{x}n_\text{y}}^\text{T}+d_{n_\text{x}n_\text{y}}^\text{R})$, we have the freedom to choose the matching value for $\phi_{n_\text{x}n_\text{y}}$.
Consequently, the optimal position of the MA corresponds to a specific point that can be determined by the relevant distances. In this scenario, the spatial repositioning of the MA does not affect the achievable gain. Thus, in an ideal scenario with continuous phase-shifters, the optimal phase of each TRIS element perfectly compensates for the propagation-induced phase term, leading to the following upper bound
\begin{align} \label{eq:UpperBound}
    & \text{SNR}_\text{UB} = 
    \frac{\lambda ^2 \Gamma^2 {P G F }}{16\pi^2 \sigma ^2} \left | \sum_{n_\text{x} = 1}^{N_\text{x}} \sum_{n_\text{y} = 1}^{N_\text{y}} \frac{1}{d_{n_\text{x}n_\text{y}}^\text{T}d_{n_\text{x}n_\text{y}}^\text{R}}  \right |^2.
\end{align}
Accordingly, the subsequent sections of this study concentrate on the practically relevant case of discrete phase-shifters, which reflects typical real-world hardware and cost constraints. 

The goal is now to maximize the SNR (\ref{eq:SNRFirst}) by optimizing the MA location and the discrete phase-shifters of the TRIS.
When employing a TRIS with discrete phase-shift adjustment, the limited resolution of the phase-shifters must be considered. Let $b$ denote the number of quantization bits. The feasible phase set is
\begin{align}
    \boldsymbol{\Phi}_b = \big \{  0, \frac{1}{2^{b-1}}\pi, \dots, \frac{2^b-1}{2^{b-1}}\pi  \big \} .   
\end{align}
We define $Q(\cdot)$ as the \emph{quantization function}, which projects the ideal phase-shift onto the nearest element of the feasible phase set $\boldsymbol{\Phi}_b$ associated with the TRIS phase-shifters:
\begin{align}
     \tilde{{\phi}}_{n_\text{x}n_\text{y}} = Q({\phi}_{n_\text{x}n_\text{y}};\boldsymbol{\Phi}_b) = \arg\min_{{\phi}_b \in \boldsymbol{\Phi}_b} \, |{\phi}_{n_\text{x}n_\text{y}} - {{\phi}_b}|.
\end{align}
As a result, we have the quantized phase-shifter vector $\tilde{\boldsymbol{\phi}} = [\tilde{\phi}_1, \tilde{\phi}_2, \dots,  \tilde{\phi}_{n_\text{x}n_\text{y}}, \dots, \tilde{\phi}_N ]$, where each element is calculated as the following
\begin{align}
    \tilde{\phi}_{n_\text{x}n_\text{y}} = Q \big( \frac{2\pi}{\lambda} (d_{n_\text{x}n_\text{y}}^\text{R}+d_{n_\text{x}n_\text{y}}^\text{T});\boldsymbol{\Phi}_b \big), 
\end{align}
and the received signal can now be described as
\begin{align}
y =  \frac{\lambda \Gamma \sqrt{P G F }}{4\pi} \sum_{n_\text{x} = 1}^{N_\text{x}} \sum_{n_\text{y} = 1}^{N_\text{y}}  \big ( \frac{1}{d_{n_\text{x}n_\text{y}}^\text{T}d_{n_\text{x}n_\text{y}}^\text{R}} e^{-j\tilde{{\phi}}_{n_\text{x}n_\text{y}}} \big )s + n_0,
\end{align}
therefore our objective would be 
\begin{align} 
    \text{SNR}_{\mathrm{Dis}} =  \frac{\lambda ^2 \Gamma^2 {P G F }}{16\pi^2 \sigma ^2} \left | \sum_{n_\text{x} = 1}^{N_\text{x}} \sum_{n_\text{y} = 1}^{N_\text{y}} \frac{1}{d_{n_\text{x}n_\text{y}}^\text{T}d_{n_\text{x}n_\text{y}}^\text{R}} e^{-j\tilde{{\phi}}_{n_\text{x}n_\text{y}}}  \right |^2.
\end{align}

To determine the optimal location for the MA, the movement region $\mathcal{R}$ can be discretized into a set of sampling points. The point that maximizes the SNR is then selected from this set, denoted as $\boldsymbol{t}_l \in \mathcal{R}$. This approach facilitates an efficient search for the configuration that yields the highest possible SNR.

The optimization problem can be formulated as 
\begin{subequations}\label{eq:series}
    \begin{align}
        \mathcal{P}_1:\ \quad & \max_{\boldsymbol{t}_l,\boldsymbol{\phi}}\ \mathrm{SNR}_{\mathrm{Dis}}, \label{eq:SNR}\\     
        \text{s.t.}  \quad & \boldsymbol{t}_l \in \mathcal{R},  \label{eq:MA.Reg}\\  
                     \quad & \lvert \tilde{\phi}_n \rvert = 1, && \forall n \in \mathcal{N} \label{eq:phiConstraint} \\
                     \quad & \arg ( \tilde{\phi}_n ) \in \boldsymbol{\Phi}_b, && \forall n \in \mathcal{N} \label{eq:MADiscrete}
    \end{align}
\end{subequations}
where \eqref{eq:MA.Reg} defines the feasible region for antenna movement.
The \eqref{eq:phiConstraint} describes the constraint of the TRIS phase-shifters, and the \eqref{eq:MADiscrete} ensures that the discrete phase-shifters are inside the defined feasible set.

However, the optimization problem (\ref{eq:SNR}) is a non-convex problem due to the discrete phase constraints and the nonlinear coupling between $\boldsymbol{t}_l$ and $\phi$.
To tackle this challenge, we employ an alternating optimization (AO) framework, which decouples the decision variables $\boldsymbol{\phi}$ and $\boldsymbol{t}_l$ in $\mathcal{P}_1$. 

The corresponding algorithm to optimize $\boldsymbol{\phi}$ and $\boldsymbol{t}_l$ is given in Algorithm \ref{alg:SimpleTRIS}.

\begin{algorithm}[H]
\caption{Alternating Optimization for Quantized TRIS}
\label{alg:SimpleTRIS}
\begin{algorithmic}[1]
\State \textbf{Inputs:} TRIS initial phase-shifter $\boldsymbol{\phi}_0$, MA initial position $\mathbf{t}_0$, $\boldsymbol{\Phi}_b$, $d_{n_\text{x}n_\text{y}}^{R}$, ${e}_{n_\text{x}n_\text{y}}$. 
\State \textbf{Initialization:} $\text{SNR}_{\max}=\!0,\ \mathbf{t}_l^\star=\!\mathbf{t}_0$
\For{$\mathbf{t}_l=(x_t,y_t,z_t)\in \mathcal{R}$}
\For{$n_{\text{x}},n_{\text{y}} \in N $}
    \State \hspace{0.3cm}${\phi}_{n_\text{x}n_\text{y}} = \tfrac{2\pi}{\lambda} (d^R_{n_\text{x}n_\text{y}} + d_{n_\text{x}n_\text{y}}^T(\mathbf{t}_l))$;
    \State \hspace{0.3cm}$\tilde{{\phi}}_{n_\text{x}n_\text{y}} = Q({\phi}_{n_\text{x}n_\text{y}};\boldsymbol{\Phi}_b)$;
    \State \hspace{0.3cm}$d_{n_\text{x}n_\text{y}}^T(\mathbf{t}_l) = \|{e}_{n_\text{x}n_\text{y}}-\mathbf{t}_l\|$;
\EndFor
\State \hspace{0.3cm}$\text{SNR}(\mathbf{t}_l) =  \left | \sum_{n_\text{x}=1}^{N_\text{x}}\sum_{n_\text{y}=1}^{N_\text{y}} \frac{1}{d^T_{n_\text{x}n_\text{y}}(\mathbf{t}_l) d^R_{n_\text{x}n_\text{y}}} e^{j(\tilde{{\phi}}_{n_\text{x}n_\text{y}})} \right |^2$
\hspace{0.3cm}\If{$\text{SNR}(\mathbf{t}_l) > \text{SNR}_{\max}$}
\hspace{0.6cm}\State $\text{SNR}_{\max}=\!\text{SNR}(\mathbf{t}_l),\quad \mathbf{t}_l^\star=\!\mathbf{t}_l$;
\hspace{0.3cm}\EndIf
\EndFor
\State \textbf{Output:} best MA position $\mathbf{t}_l^\star$, $\text{SNR}_{\max}$.
\end{algorithmic}
\end{algorithm}

This algorithm uses a search-based method defined by a finite number of sampling points and a limited number of iterations, ensuring theoretical convergence. It should also be noted that, for larger TRIS sizes and more sampling points in the antenna region, the convergence time increases accordingly.

\section{Simulation Results}
In this section, simulations are presented to illustrate the effects of employing a MA under a discrete phase-shifter TRIS condition.
We assume the carrier frequency $f = 20\,\mathrm{GHz}$ and TRIS element spacing of $\lambda/2$ (which would be $\sim 1.5\,\mathrm{cm}$ at $20\,\mathrm{GHz}$). We also consider the transmit power $P = 13.6\,\mathrm{dBm}$ and the noise variance $\sigma^2 = -120\,\mathrm{dBm}$. The user is located at a distance of $12.5\, \mathrm{m}$ of the TRIS. 

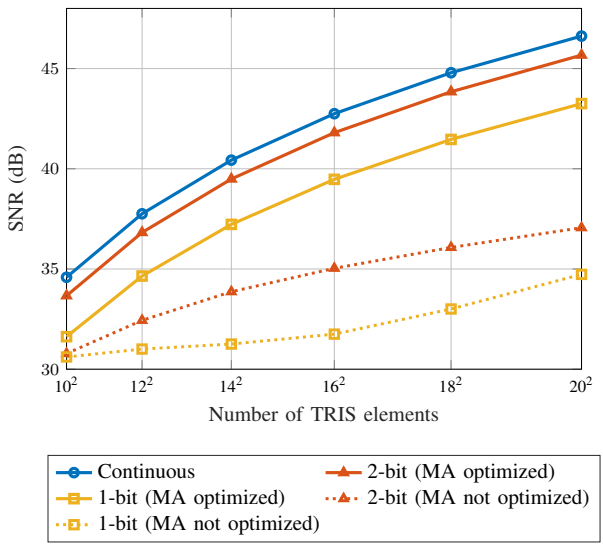
\begin{figure}[ht]
  \centering
%
%
\definecolor{mycolor1}{rgb}{0.00000,0.44700,0.74100}%
\definecolor{mycolor2}{rgb}{0.85000,0.32500,0.09800}%
\definecolor{mycolor3}{rgb}{0.92900,0.69400,0.12500}%
\definecolor{mycolor4}{rgb}{0.49400,0.18400,0.55600}%
\definecolor{mycolor5}{rgb}{0.46600,0.67400,0.18800}%
\begin{tikzpicture}[scale = 0.8, transform shape]

\begin{axis}[%
width=8.558cm,
height=6cm,
at={(0cm,0cm)},
scale only axis,
xmin=100,
xmax=400,
xtick={100,144,196,256,324,400},
xticklabels={{$\text{10}^{\text{2}}$},{$\text{12}^{\text{2}}$},{$\text{14}^{\text{2}}$},{$\text{16}^{\text{2}}$},{$\text{18}^{\text{2}}$},{$\text{20}^{\text{2}}$}},
xlabel style={font=\color{white!15!black}},
xlabel={Number of TRIS elements},
ymin=30,
ymax=48,
ylabel style={font=\color{white!15!black}},
ylabel={SNR (dB)},
axis background/.style={fill=white},
title style={font=\bfseries},
xmajorgrids,
ymajorgrids,
legend style={
  at={(0.5,-0.48)},
  anchor=south,         
  draw=white!15!black,  
  legend cell align=left,
  align=left,
  font=\normalsize,     
  legend columns=2      
}
]
\addplot [color=mycolor1, line width=1.5pt, mark=o, mark options={solid, mycolor1}]
  table[row sep=crcr]{%
100	34.5880504160247\\
144	37.7539988423128\\
196	40.4303331553297\\
256	42.74823828659\\
324	44.7923313995894\\
400	46.6203886958912\\
};
\addlegendentry{Continuous}

\addplot [color=mycolor2, line width=1.5pt, mark=triangle, mark options={solid, mycolor2}]
  table[row sep=crcr]{%
100	33.6649488662626\\
144	36.820541204803\\
196	39.48943203955\\
256	41.8011473591501\\
324	43.842329094998\\
400	45.6697482106596\\
};
\addlegendentry{2-bit (MA optimized)}

\addplot [color=mycolor3, line width=1.5pt, mark=square, mark options={solid, mycolor3}]
  table[row sep=crcr]{%
100	31.6232712599596\\
144	34.6483764381731\\
196	37.2262599152259\\
256	39.4729421716871\\
324	41.4641548647106\\
400	43.251206983257\\
};
\addlegendentry{1-bit (MA optimized)}

\addplot [color=mycolor2, dotted, line width=1.4pt, mark=triangle, mark options={solid, mycolor2}]
  table[row sep=crcr]{%
100	30.7803805854983\\
144	32.4370551560182\\
196	33.864916750293\\
256	35.0353892310814\\
324	36.0750045554175\\
400	37.059002033749\\
};
\addlegendentry{2-bit (MA not optimized)}

\addplot [color=mycolor3, dotted, line width=1.4pt, mark=square, mark options={solid, mycolor3}]
  table[row sep=crcr]{%
100	30.6107499639835\\
144	31.0056730946166\\
196	31.2551062444963\\
256	31.7497281349745\\
324	33.0050232341671\\
400	34.7317536654271\\
};
\addlegendentry{1-bit (MA not optimized)}

\end{axis}
\end{tikzpicture}%
  \caption{Received SNR vs. TRIS size for different quantization bits for the phase-shifters, with and without MA optimization.}
  \label{fig:snr_size}
\end{figure}
Fig.~\ref{fig:snr_size} illustrates the received SNR as a function of the TRIS size across a range of configuration scenarios, including both optimized and non-optimized MA designs. The scenario featuring continuous phase control is depicted as the theoretical upper bound of performance according to each TRIS size. Increasing the phase quantization resolution, measured in bits, substantially improves the achievable SNR at the receiver. This enhancement significantly reduces the performance gap compared to the continuous case. These results underscore the practical benefits of even modest enhancements in phase resolution for real-world deployments. Moreover, the figure shows that MA location optimization further augments system performance across all TRIS configurations. 

\begin{figure}[ht]
  \centering
  %

\definecolor{mycolor1}{rgb}{0.00000,0.44700,0.74100}%
\definecolor{mycolor2}{rgb}{0.85000,0.32500,0.09800}%
\definecolor{mycolor3}{rgb}{0.92900,0.69400,0.12500}%
\begin{tikzpicture}[scale = 0.8]

\begin{axis}[%
width=8.558cm,
height=6cm,
at={(0cm,0cm)},
scale only axis,
xmin=1,
xmax=2.8,
xlabel style={font=\color{white!15!black}},
xlabel={MA-TRIS distance [m]},
ymin=33,
ymax=43,
ylabel style={font=\color{white!15!black}},
ylabel={SNR (dB)},
axis background/.style={fill=white},
title style={font=\bfseries},
xmajorgrids,
ymajorgrids,
legend style={legend cell align=left, align=left, draw=white!15!black}
]
\addplot [color=mycolor1, line width=1.5pt, mark=o, mark options={solid, mycolor1}]
  table[row sep=crcr]{%
1	42.8054860659794\\
1.15909090909091	41.5269758112419\\
1.31818181818182	40.4123586280311\\
1.47727272727273	39.4244170535221\\
1.63636363636364	38.537311356661\\
1.79545454545455	37.7323728626955\\
1.95454545454545	36.9956761636357\\
2.11363636363636	36.3165582331576\\
2.27272727272727	35.686673463918\\
2.43181818181818	35.0993676585031\\
2.59090909090909	34.5492499841296\\
2.75	34.0318922514266\\
};
\addlegendentry{Continuous}

\addplot [color=mycolor2, line width=1.5pt, mark=square, mark options={solid, mycolor2}]
  table[row sep=crcr]{%
1	41.4398830218122\\
1.15909090909091	40.4574884519061\\
1.31818181818182	39.5602225932918\\
1.47727272727273	38.7329620871576\\
1.63636363636364	37.9661102081588\\
1.79545454545455	37.2525440936938\\
1.95454545454545	36.5865060614224\\
2.11363636363636	35.9630648728968\\
2.27272727272727	35.3778393695539\\
2.43181818181818	34.8269129835431\\
2.59090909090909	34.3068316515995\\
2.75	33.8145684978058\\
};
\addlegendentry{1-bit}

\addplot [color=mycolor3, line width=1.5pt, mark=triangle, mark options={solid, mycolor3}]
  table[row sep=crcr]{%
1	42.1845635925677\\
1.15909090909091	40.9341934006224\\
1.31818181818182	39.8521494299522\\
1.47727272727273	38.9059289472097\\
1.63636363636364	38.0687940224864\\
1.79545454545455	37.3145859445645\\
1.95454545454545	36.6236523821969\\
2.11363636363636	35.9844888026475\\
2.27272727272727	35.3896201191873\\
2.43181818181818	34.8334026904268\\
2.59090909090909	34.310551686744\\
2.75	33.8167688953868\\
};
\addlegendentry{2-bit}

\addplot [color=black, dashed, line width=1.2pt]
  table[row sep=crcr]{%
1.9186717312	33\\
1.9186717312	43\\
};
\addlegendentry{Near-field / Far-field boundary}

\end{axis}
\end{tikzpicture}%
  \caption{SNR vs. MA-TRIS Distance for an optimized MA.}
  \label{fig:snr_MA_Distance}
\end{figure}
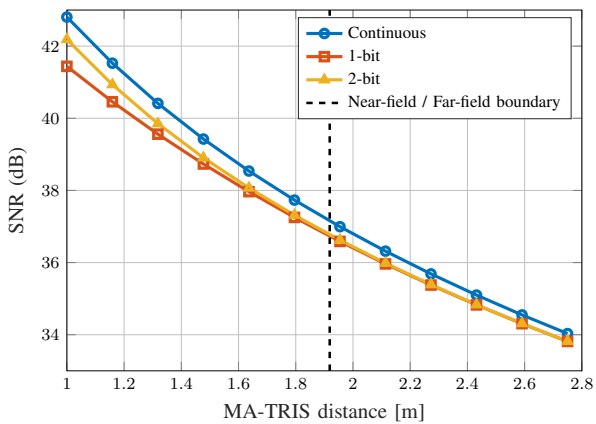
Building on the previous results, Fig. \ref{fig:snr_MA_Distance} shows the dependence of the received SNR on the distance separating the MA and the TRIS. Across all evaluated configurations, the SNR demonstrates a monotonic decline as the separation increases, with the continuous phase configuration establishing the upper bound, and the 2-bit and 1-bit quantized cases exhibiting slightly reduced SNR values. The vertical dashed line in the figure indicates the boundary between near-field and far-field regions at $D_\text{Ray}\approx1.91\,$m, corresponding to a TRIS size of $N = 16^2$. Another noteworthy observation is that, in the near-field region, the positive impact of increasing the number of quantization bits on the SNR is clearly evident. In contrast, in the far-field region, this difference is less pronounced.

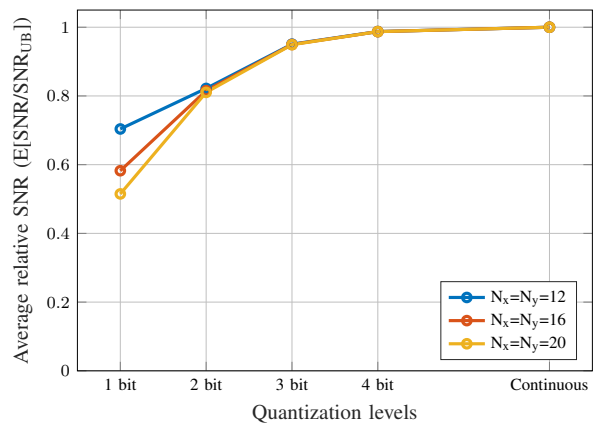
\begin{figure}[ht]
  \centering
%

\definecolor{mycolor1}{rgb}{0.00000,0.44700,0.74100}%
\definecolor{mycolor2}{rgb}{0.85000,0.32500,0.09800}%
\definecolor{mycolor3}{rgb}{0.92900,0.69400,0.12500}%
\definecolor{mycolor4}{rgb}{0.49400,0.18400,0.55600}%
\definecolor{mycolor5}{rgb}{0.46600,0.67400,0.18800}%
\definecolor{mycolor6}{rgb}{0.30100,0.74500,0.93300}%
\begin{tikzpicture}[scale = 0.8]

\begin{axis}[%
width=8.558cm,
height=6cm,
at={(0cm,0cm)},
scale only axis,
xmin=0.5,
xmax=6.5,
xtick={1,2,3,4,6},
xticklabels={{1 bit},{2 bit},{3 bit},{4 bit},{Continuous}},
xlabel style={font=\color{white!15!black}},
xlabel={Quantization levels},
ymin=0,
ymax=1.05,
ylabel style={font=\color{white!15!black}},
ylabel={$\text{Average relative SNR    (E[SNR}\text{/SNR}_{\text{UB}}\text{]}$)  },
axis background/.style={fill=white},
title style={font=\bfseries},
xmajorgrids,
ymajorgrids,
legend style={at={(0.97,0.03)}, anchor=south east, legend cell align=left, align=left, draw=white!15!black}
]


\addplot [color=mycolor1, line width=1.5pt, mark=o, mark options={solid, mycolor1}]
  table[row sep=crcr]{%
1	0.703866022417854\\
2	0.821859086313974\\
3	0.950714069445942\\
4	0.987378717043764\\
6	1\\
};
\addlegendentry{$\text{N}_\text{x}\text{=N}_\text{y}\text{=12}$}


\addplot [color=mycolor2, line width=1.5pt, mark=o, mark options={solid, mycolor2}]
  table[row sep=crcr]{%
1	0.582382514763685\\
2	0.815290179008079\\
3	0.949803382752237\\
4	0.987236372548947\\
6	1\\
};
\addlegendentry{$\text{N}_\text{x}\text{=N}_\text{y}\text{=16}$}


\addplot [color=mycolor3, line width=1.5pt, mark=o, mark options={solid, mycolor3}]
  table[row sep=crcr]{%
1	0.514876685389454\\
2	0.81035053472799\\
3	0.949127027853493\\
4	0.987187497189255\\
6	1\\
};
\addlegendentry{$\text{N}_\text{x}\text{=N}_\text{y}\text{=20}$}

\end{axis}
\end{tikzpicture}%
  \caption{Average relative SNR vs. Quantization levels for an optimized MA.}
  \label{fig:gain_bit}
\end{figure}
Finally, Fig. \ref{fig:gain_bit} demonstrates the remarkable adaptability of the TRIS, even when subject to constraints imposed by limited discrete phase values. The figure systematically quantifies performance loss as a function of quantization bit depth and array size. Each curve corresponds to a specific square TRIS configuration (e.g., $N_x = N_y = 12, 16, 20$), with the horizontal axis representing phase resolution (1–4 bits) and including the ideal continuous scenario for reference. The vertical axis presents the \emph{average relative SNR}, computed over the movement region of the transmit antenna. At each grid location, the SNR achieved under $b$-bit quantization is normalized to the SNR in the corresponding continuous case, $\text{SNR}_\text{UB}$, at that position; these ratios are then averaged. Values approaching unity indicate minimal loss compared to the ideal, while lower values correspond to increased performance degradation. The plot highlights two key trends: (i) Increasing the number of quantization bits consistently mitigates performance loss, bringing the curves progressively closer to the ideal; and (ii) variations among TRIS array sizes manifest as subtle differences in the curves, indicating that system performance sensitivity to phase quantization is also influenced by array size. Since user location, antenna movement region, and sampling grid remain constant across all scenarios, the observed differences can be attributed directly to TRIS size and quantization level, rather than to changes in geometry or measurement conditions.

\section{Conclusion}
This paper introduces a compact BS design that combines a MA with a TRIS in the near field. We provide an upper bound and develop an efficient method to jointly optimize MA position and TRIS phases, even with discrete-phase shifters. Our results show that TRIS with continuous phase-shift adjustment makes MA placement less critical, while TRIS under discrete phase-shift setting benefits significantly from MA repositioning, which improves SNR. Scaling the TRIS aperture also enhances performance, approaching the ideal continuous-phase case. Integrating discrete TRIS hardware with MA control enables a practical, energy-efficient transmitter that retains most benefits of ideal phase control, highlighting the scalability and versatility of the proposed design.

\FloatBarrier
\bibliographystyle{IEEEtran}
\bibliography{Bib.bib}
\end{document}